\documentclass[aps,prl,twocolumn,superscriptaddress]{revtex4-2}
\usepackage{bbm}
\usepackage{mathrsfs}
\usepackage{amsmath}
\usepackage{amsfonts}
\usepackage[colorlinks=true,linkcolor=blue,urlcolor=blue,citecolor=blue,anchorcolor=blue]{hyperref}
\usepackage{graphicx,epstopdf}
\usepackage{subfigure}
\usepackage{epsfig}
\usepackage{dcolumn}
\usepackage{bm}
\usepackage{color}
\usepackage{natbib}
\usepackage{amssymb}
\usepackage{xcolor}
\usepackage{braket}
\usepackage{ulem}
\usepackage{float}
\usepackage{lipsum}
\usepackage{tikz}

\definecolor{newtxtcolor1}{rgb}{0.0, 0, 1.0}

\definecolor{newtxtcolor2}{rgb}{0.3, 0.36, 0.77}

\begin{document}

\title{Characterizing the Multipartite Entanglement Structure of Non-Gaussian Continuous-Variable States with a Single Evolution Operator}

 \author{Mingsheng Tian}
 \thanks{These authors contributed equally to this work.}

 \address{State Key Laboratory for Mesoscopic Physics, School of Physics, Frontiers Science Center for Nano-optoelectronics, $\&$ Collaborative Innovation Center of Quantum Matter, Peking University, Beijing 100871, China}
  \address{Department of Physics, The Pennsylvania State University, University Park, Pennsylvania, 16802, USA}
 
 \author{Xiaoting Gao}
 \thanks{These authors contributed equally to this work.}
   \address{State Key Laboratory for Mesoscopic Physics, School of Physics, Frontiers Science Center for Nano-optoelectronics, $\&$ Collaborative Innovation Center of Quantum Matter, Peking University, Beijing 100871, China}

 \author{Boxuan Jing}
 \address{State Key Laboratory for Mesoscopic Physics, School of Physics, Frontiers Science Center for Nano-optoelectronics, $\&$ Collaborative Innovation Center of Quantum Matter, Peking University, Beijing 100871, China}

 \author{Fengxiao Sun}
   \address{State Key Laboratory for Mesoscopic Physics, School of Physics, Frontiers Science Center for Nano-optoelectronics, $\&$ Collaborative Innovation Center of Quantum Matter, Peking University, Beijing 100871, China}

\author{Matteo Fadel}
\affiliation{Department of Physics, ETH Z\"{urich}, 8093 Z\"{urich}, Switzerland}

\author{Manuel Gessner}
\affiliation{Instituto de Física Corpuscular (IFIC), CSIC‐Universitat de València and Departament de Física Teòrica, UV, C/Dr Moliner 50, E-46100 Burjassot (Valencia), Spain}

\author{Qiongyi He}
\email{qiongyihe@pku.edu.cn}
\address{State Key Laboratory for Mesoscopic Physics, School of Physics, Frontiers Science Center for Nano-optoelectronics, $\&$ Collaborative Innovation Center of Quantum Matter, Peking University, Beijing 100871, China}
\address{Collaborative Innovation Center of Extreme Optics, Shanxi University, Taiyuan, Shanxi 030006, China}
\address{Peking University Yangtze Delta Institute of Optoelectronics, Nantong, Jiangsu 226010, China}
\address{Hefei National Laboratory, Hefei 230088, China}

\begin{abstract}
Multipartite entanglement is an essential resource for quantum information tasks, but characterizing entanglement structures in continuous variable systems remains challenging, especially in multimode non-Gaussian scenarios. In this work, we introduce an efficient method for detecting multipartite entanglement structures in continuous-variable states. Based on the quantum Fisher information, we propose a systematic approach to identify an encoding operator that can efficiently capture the quantum correlations in multimode non-Gaussian states. We demonstrate the effectiveness of our method on over $10^5$ randomly generated multimode-entangled quantum states, achieving a very high success rate in entanglement detection. Additionally, the robustness of our method can be considerably enhanced against losses by expanding the set of accessible operators. This work provides a general framework for characterizing entanglement structures in diverse continuous variable systems, enabling a number of experimentally relevant applications.
\end{abstract}

\maketitle
Multipartite entanglement is an essential resource for many quantum computation, sensing and communication protocols~\cite{vanloock2000,gisin2002-rmp-gme-qc,otfried2009,bancal2011-gme-qc,das2021-gme-qc,giovannetti2004-gme-metrology,briegel2009-gme-comp,bao2023very,yokoyama2013ultra,FrerotROPP2023}. Its importance is particularly evident in continuous variable (CV) systems, where multimode Gaussian states, such as squeezed states and cluster states, can be deterministically prepared and controlled~\cite{yokoyama2013-cv,andersen2015-cv,armstrong2015-cv}. 
However, to achieve an advantage over classical protocols in CV systems, non-Gaussian features are necessary~\cite{mattia2021-prxquantum-ng}. Non-Gaussian states play a crucial role for quantum computation~\cite{niset2009-prl-ng-comp,mari2012-prl-ng-comp,chabaud2023-prl-ng-comp}, sensing~\cite{joo2011-prl-ng-qm,strobel2014-science-ng-qm,hanamura2023-ng}, and imaging~\cite{liu2021-ghost,karuseichyk2022-prr-ng-imag}. Strategies to generate and control such states has seen substantial advancements in recent years, including through photon-subtraction~\cite{namekate2010-np,nicolas2020np} and nonlinear operations, such as spontaneous parametric down-conversion (SPDC)~\cite{prxthreemode,spdc2004} and Kerr interactions in microwave cavities coupled superconducting artificial atoms~\cite{wang2016schrodinger,matteo2023-science,matteo2024-np}.

Despite these advancements, accurately characterizing both Gaussian and non-Gaussian multimode entanglement in CV systems continues to present significant challenges~\cite{mattia2021-prxquantum-ng}, e.g. due to the complex correlations appearing among higher-order statistical moments of quadrature field operations. Unlike discrete variable (DV) systems, where entanglement structures are identified based on separable partitions and their sizes~\cite{hyllus2012-pra-fisher,PhysRevA.85.022322,manuel2021prl-young,Fadel_2023,pan2020-prx}, there is no similarly comprehensive method for CV systems. State-of-the-art techniques in CV systems predominantly analyze bi- and multipartite entanglement through first and second-order moments~\cite{duan2000,simon2000,van-loock2003,threecolor,reidpra2014,PhysRevLett.117.110502,weekbrook2012-rmp,TehPRA22}. Since non-Gaussian states involve more complex higher-order correlations, these techniques are mostly limited to the study of Gaussian CV entanglement. Criteria for detecting non-Gaussian CV entanglement~\cite{PhysRevLett.111.110503,pra2016manuel,Gessner2017entanglement,suxiaolong-npj2019,Fadel2022entanglementoflocal,Barral_2024,Tian_2024} can only distinguish entanglement properties of individual partitions and need to use different operators to capture the correlations for other partitions, which become experimentally ineffective to characterize entanglement structure. This limitation also hinders the understanding of complex structures. Therefore, there is a pressing need to develop an experimentally friendly method to characterize the multipartite entanglement structure in non-Gaussian CV systems, filling the gap in the field.

Here,  we propose a general method to characterize entanglement structure based on quantum metrology tools.
Specifically, we systematically determine an operator that can effectively capture different correlations in multimode systems.
Our approach begins with the analytical identification of a convex set of operators, each capable of witnessing specific bi-inseparable entanglement structures across arbitrary quantum states. Since different entanglement structures correlate with distinct convex sets, we then proceed to pinpoint the intersection of these sets. This intersection allows us to identify an operator for efficiently witnessing entanglement. 
Crucially, contrary to previous work where one needs to choose different operators to identify different quantum correlations, our method identifies various quantum correlations with the same operator, which enables us to efficiently characterize and detect entanglement structures.
To demonstrate the practicality and efficiency of our method, we apply it to a broad range of non-Gaussian states generated through various stellar ranks and nonlinear processes, including over $10^5$ randomly produced quantum states. Furthermore, we examine the robustness of our method in scenarios involving loss, demonstrating that the inclusion of high-order operators significantly enhances resilience to loss channels. 
This approach marks the accessible method for witnessing entanglement structure in arbitrary  Gaussian and non-Gaussian states, thereby addressing a significant need in CV systems.

\begin{figure}
    \centering
    \includegraphics[width=8.5cm]{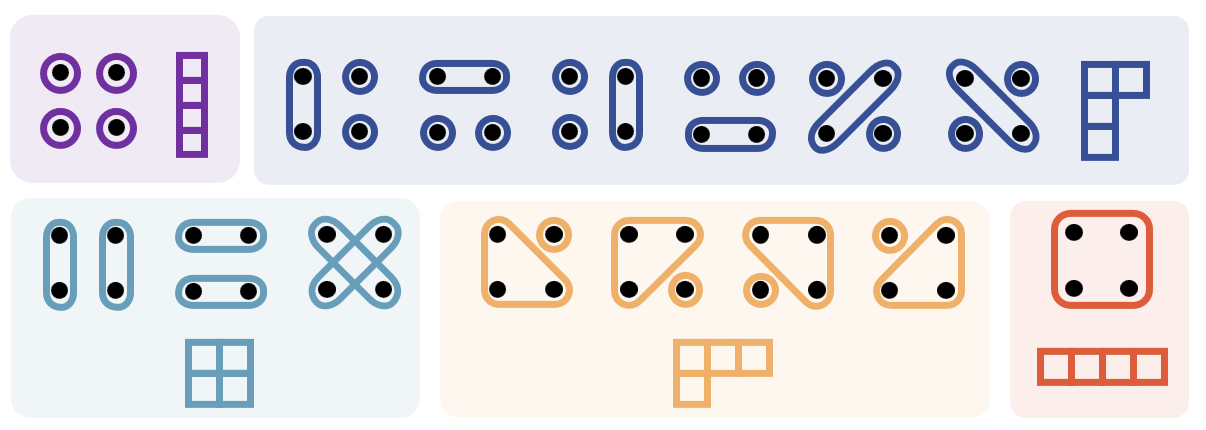}
    \caption{Graph representation of different multipartite entanglement structures. As illustrated, a system with $N=4$ modes can be 
    characterized by 15 different partitions, associated to different distributions of correlations as indicated by the solid circles.
    Each partition can be graphically represented by a Young tableau, where boxes in each row indicate correlated modes, while vertically stacked boxes indicate separable subsets.
    }
\label{fig1}
\end{figure}


\vspace{2mm}
\textit{Metrological bound for $\mathcal{K}$-partitions}.--
We begin discussing the entanglement structure for $N$ parties. A partition $\mathcal{K}=\left\{H_1, H_2, \ldots, H_{k}\right\}$ separates the total $N$-partite system into $k$ nonempty, disjoint subsets $H_l$, where each subset has a size $N_l$ such that $\sum_{l=1}^{k} N_l=N$. A state $\hat{\rho}_{\mathcal{K}}$ is $\mathcal{K}$-separable if there exist local quantum states $\hat{\rho}_{H_l}^{(\gamma)}$ for each subset and a probability distribution $p_\gamma$ such that $\hat{\rho}_{\mathcal{K}}=\sum_\gamma p_\gamma \hat{\rho}_{H_1}^{(\gamma)} \otimes \cdots \otimes \hat{\rho}_{H_{k}}^{(\gamma)}$. A partition $\mathcal{K}$ can be classified in terms of the number and respective sizes of the separable subsets. This can be illustrated graphically by means of a Young tableau, as illustrated in Fig.~\ref{fig1} for a 4-mode system.

Using the framework of quantum metrology, it is known that for an arbitrary $\mathcal{K}$-separable state $\hat{\rho}_{\mathcal{K}}$, 
the quantum Fisher information (QFI) is bounded by the variances as~\cite{pra2016manuel,qfi2018-rmp}
\begin{equation}\label{eqfv}
F_Q(\hat{\rho}_{\mathcal{K}},   \hat{A}) \leq 
4 \mathcal{V}(\hat{\rho}_{\mathcal{K}},  \hat{A}).
\end{equation}
Here, $\hat{A}= \sum_{j=1}^N \hat{A}_j$, where $\hat{A}_j$ represents a local generator in the reduced state $\hat{\rho}_{H_j}$. 
$ \mathcal{V}(\hat{\rho}_{\mathcal{K}}, \hat{A})=\sum_{i=1}^k\mathrm{Var}(\hat{\rho}_{H_i},\sum_{j\in H_i}\hat{A}_{j})$ defines the total variance, where
$\mathrm{Var}(\hat{\rho},\hat{O}) = \langle \hat{O}^2 \rangle_{\hat{\rho}} - \langle \hat{O} \rangle^2_{\hat{\rho}}$,  and $\hat{\rho}_{H_i}$ is the reduced state of subsystem $H_i$. 
The explicit expression of the QFI is given by
$
F_Q[\hat{\rho}_{\mathcal{K}},\hat{A}]
=2 \sum_{\substack{k, l}} \frac{(p_{k}-p_{l})^{2}}{p_{k}+p_{l}} | \langle\Psi_{k}|\hat{A}| \Psi_{l}\rangle|^{2}
$, 
with the spectral decomposition $\hat{\rho}=\sum_k p_k|\Psi_k\rangle \langle \Psi_k|$~\cite{qfi1994}. 
Since Eq.~(\ref{eqfv}) is a necessary criterion for separability, its violation implies $\mathcal{K}$-inseparability.

The right-hand side of the criterion~(\ref{eqfv}) is tailored to detect inseparability within the specific partition $\mathcal{K}$. 
To maximize entanglement detection, one must optimize the operator $\hat{A}$ to achieve the largest possible violation of Eq.~(\ref{eqfv}). Typically, a different operator must be chosen in order to efficiently witness the entanglement across a different partition. Therefore, if an operator $\hat{A}$ exists for a quantum state $\hat{\rho}$ that can capture correlations across all partitions corresponding to a given Young diagram, as illustrated in Fig.~\ref{fig1}, it offers a more efficient way to characterize the entanglement structure.
To this aim, we propose a general and systematic optimization method to characterize different partitions with the same operator $\hat{A}$. The methodology consists of the following steps:

\vspace{2mm}
\textit{Step 1.--- Finding the convex sets of operators witnessing entanglement for each $\mathcal{K}$-partition.}

To witness $\mathcal{K}$-partite entanglement through inequality~(\ref{eqfv}), the choice of suitable local operators $\hat{A}_j$ is crucial and dependent on the given state $\hat{\rho}$. 
These operators can be constructed by analytically optimizing over linear combinations of accessible operators forming the set $\mathbf{S}_j = \{\hat{S}_j^{(1)}, \hat{S}_j^{(2)}, \cdots\}$~\cite{pra2016manuel}. Specifically, we express $\hat{A}_j$ as the linear combination
\begin{equation}
    \hat{A}_j = \sum_{m=1}^L c_j^{(m)} \hat{S}_j^{(m)} = \mathbf{c}_j \cdot \mathbf{S}_j  \;,
\end{equation}
 {where $\mathbf{c}_j=(c_j^{(1)},c_j^{(2)},...,c_j^{(L)})$ is the vector of coefficients.
}
A typical choice for such a set includes local position operators $\hat{x}_j$ and momentum operators $\hat{p}_j$. However, since the states we consider are in general non-Gaussian, simply measuring linear observables is insufficient for characterizing their correlations. We therefore extend the family of accessible operators by incorporating higher-order moments of local quadrature operators, e.g.
$
\mathbf{S}_j=\left(\hat{x}_j,\hat{p}_j,\hat{x}_j^2,\hat{p}_j^2,(\hat{x}_j\hat{p}_j+\hat{p}_j\hat{x}_j)/2\right)
$.
 {
The full operator
$
\hat{A}(\mathbf{c}) = \sum_{j=1}^{N} \hat{A}_j = \sum_{j=1}^{N} \mathbf{c}_j \cdot \mathbf{S}_j
$
is characterized by the combined vector 
$
\mathbf{c}=(\mathbf{c}_1,\cdots,\mathbf{c}_N)^T
$
which contains $N\times L$ elements.
}
According to Eq.~(\ref{eqfv}), the quantity 
$
W = F_Q[\hat{\rho}_{\mathcal{K}},   \hat{A}] -
4 \mathcal{V}(\hat{\rho}_{\mathcal{K}},  \hat{A})
$
must be nonpositive 
when the state is separable. To witness entanglement, we maximize $W$ by varying $\mathbf{c}$ to obtain an optimal operator $\hat{A}(\mathbf{c}^{\text{opt}})$. This optimization problem can be formulated mathematically by expressing $W$ in a Rayleigh quotient form
$
W(\mathcal{M}, \mathbf{c}) = \frac{\mathbf{c}^T \mathcal{M} \mathbf{c}}{\mathbf{c}^T \mathbf{c}}
$.
 {
$\mathcal{M}$ is a Hermitian matrix constructed from the operator set and density matrix, and is given by  
$
\mathcal{M} = Q_{\hat{\rho}}^{\mathcal{S}} - 4 \Gamma_{\Pi(\hat{\rho})}^{\mathcal{S}}.
$
Both $Q_{\hat{\rho}}^{\mathcal{S}}$ and $\Gamma_{\Pi(\hat{\rho})}^{\mathcal{S}}$ are $NL \times NL$ matrices.  
Using the spectral decomposition $\hat{\rho} = \sum_k p_k |\Psi_k\rangle \langle\Psi_k|$, the elements of $Q_{\hat{\rho}}^{\mathcal{S}}$ are defined as  
$
(Q_{\hat{\rho}}^{\mathcal{S}})_{ij}^{mn} = 2 \sum_{k,l} \frac{(p_k - p_l)^2}{p_k + p_l} \langle \Psi_k | \hat{S}_i^{(m)} | \Psi_l \rangle \langle \Psi_l | \hat{S}_j^{(n)} | \Psi_k \rangle.
$
The indices $(i, j, m, n)$ specify the position of each element in the matrix, with [row, column] = [$(i - 1)L + m$, $(j - 1)L + n$]. Here $i$ and $j$ label different parties, while $m$ and $n$ label the local operators in each party.
Similarly, the elements of the covariance matrix $\Gamma_{\Pi(\hat{\rho})}^{\mathcal{S}}$ are expressed as $(\Gamma_{\hat{\rho}}^{\mathcal{{S}}})_{i j}^{m n}=\mathrm{Cov}(\hat{{S}}_{i}^{(m)}, \hat{{S}}_{j}^{(n)})_{\hat{\rho}}$ (see the Supplemental Material for details~\cite{supplementary}).
}

To determine the convex sets of operators for $\mathcal{K}$-partition inseparability, we decompose the matrix $\mathcal{M}$ as follows:
$
\mathcal{M} = U D U^T,
$
where $U$ is an orthogonal matrix, satisfying $U^T U = U U^T = I$, and $D$ is a diagonal matrix containing the eigenvalues $\lambda_i$ of $\mathcal{M}$. A normalized vector $\mathbf{c}$ can be expressed by the eigenvectors of $\mathcal{M}$, i.e.,
$
\mathbf{c} = \sum_i n_i \mathbf{e}_i.
$
Substituting this into the Rayleigh quotient $W$, we obtain:
\begin{equation}
        W(\mathcal{M},\mathbf{c})= 
    \sum_{ij}n_i n_j D_{ij} = \sum_i n_i^2 \lambda_i.
\end{equation}
This indicates that if there is a subset of eigenvectors $ Q $, where  $ \forall\mathbf{e}_i \in Q $, $ W(\mathcal{M}, \mathbf{e}_i) > 0 $, then any linear combination $ \mathbf{c} = \sum_i n_i \mathbf{e}_i $ will also satisfy $ W(\mathcal{M}, \mathbf{c}) > 0 $. Therefore, all vectors $ \mathbf{e}_i $ for which $ W(\mathcal{M}, \mathbf{e}_i) > 0 $ form convex sets. This completes \textit{Step 1}.

\vspace{2mm}
\textit{Step 2.---Determining the intersection of convex sets for different types of entanglement structures}.

For each $\mathcal{K}$-partition of the $N$ parties, the convex sets of operators determined in \textit{Step 1} corresponds to a subspace.
Different $\mathcal{K}$-partitions will in general generate different subspaces $Q_j$, and the overlap of these subspaces $P=\cap_{j=1}^{m} Q_j
$ (if any) indicates a common set of operators that allows us to certify different $\mathcal{K}$-partition inseparability.

\vspace{2mm}
\textit{
Step 3.---Find the suitable operator to witness entanglement within the intersection space.}
Once we have determined the intersection of the convex sets, any operator within this intersection can potentially detect the entanglement structure. To find a suitable one, we define a parameter
$g$ as follows:
\begin{equation}\tag{4}\label{eqg}
    g = \min\{W(\mathcal{M}_1, \mathbf{c}), W(\mathcal{M}_2, \mathbf{c}), W(\mathcal{M}_3, \mathbf{c}), \ldots\},
\end{equation}
where each $\mathcal{M}_l$ is associated to the corresponding $\mathcal{K}$-partition.
 {
Ideally, we would like to maximize $g$, but this is typically challenging because it requires a global optimization over all components of $\mathbf{c}$
(the number of parameters scales as $M^2 N$ for an $N$-mode system using $M$-th order quadratures).
}

To make the problem more tractable, we define a combined function by summing over all $\mathcal{M}_l$:  
\begin{equation}\tag{5}\label{eq555}
        f = \sum_l W(\mathcal{M}_l,\mathbf{c}) = W\left(\sum_l \mathcal{M}_l,\mathbf{c}\right).
\end{equation}
We then select the positive eigenvalues $\{\lambda_j\}$ of $\sum_l \mathcal{M}_l$. 
 {For each selected $\lambda_j$, we examine its corresponding eigenvector to identify which $\mathbf{c}$ yields the largest $g$.
Specifically, suppose that we write
$\mathbf{c} = \sum_i n_i\,\mathbf{p}_i$,
where $\{\mathbf{p}_i\}$ are the bases for the intersection space $P$, and $\{n_i\}$ are the coefficients. Substituting this into Eq.~(\ref{eq555}) gives  
$
f = \sum_{i,j} n_i \bigl(\mathbf{p}_i^T\,\mathcal{M}\,\mathbf{p}_j\bigr)\,n_j 
   \;=\; \mathbf{n}^T \,\mathcal{Q}\,\mathbf{n}
$.
We then identify the eigenvectors ${\mathbf{n}^{(j)}}$ of $\mathcal{Q}$ with positive eigenvalues ${\lambda_j}$. For each, we construct $\mathbf{c} = \sum_i n_i \mathbf{p}_i$, evaluate $g$, and select those vectors $\mathbf{c}$ that yield the largest $g$.
While this approach does not necessarily find the global maximum of $g$, it provides a practical and solvable solution: instead of directly optimizing $M^2 N$ continuous parameters (indeed it can be NP-hard~\cite{hillar2013}), we only need to solve an eigenvalue problem for an $(M^2 N)\times(M^2 N)$ matrix, which is easier because it can be done in guaranteed polynomial time with numerical linear algebra algorithms.
In addition, if the intersection space $P$ is not found from \textit{Steps} 1 and 2, one can choose a larger Hilbert space in Eq.~(\ref{eq555}) to determine $\mathbf{c}$, serving as a complement to \textit{Steps} 1 and 2.
}

These three steps provide a general method for finding an operator to witness entanglement for different partitions. 
In the following, we apply this methodology to various general CV states (both Gaussian and non-Gaussian), demonstrating its effectiveness in characterizing multipartite entanglement structures.

\begin{figure}[tb]
    \centering
    \includegraphics[width=8cm]{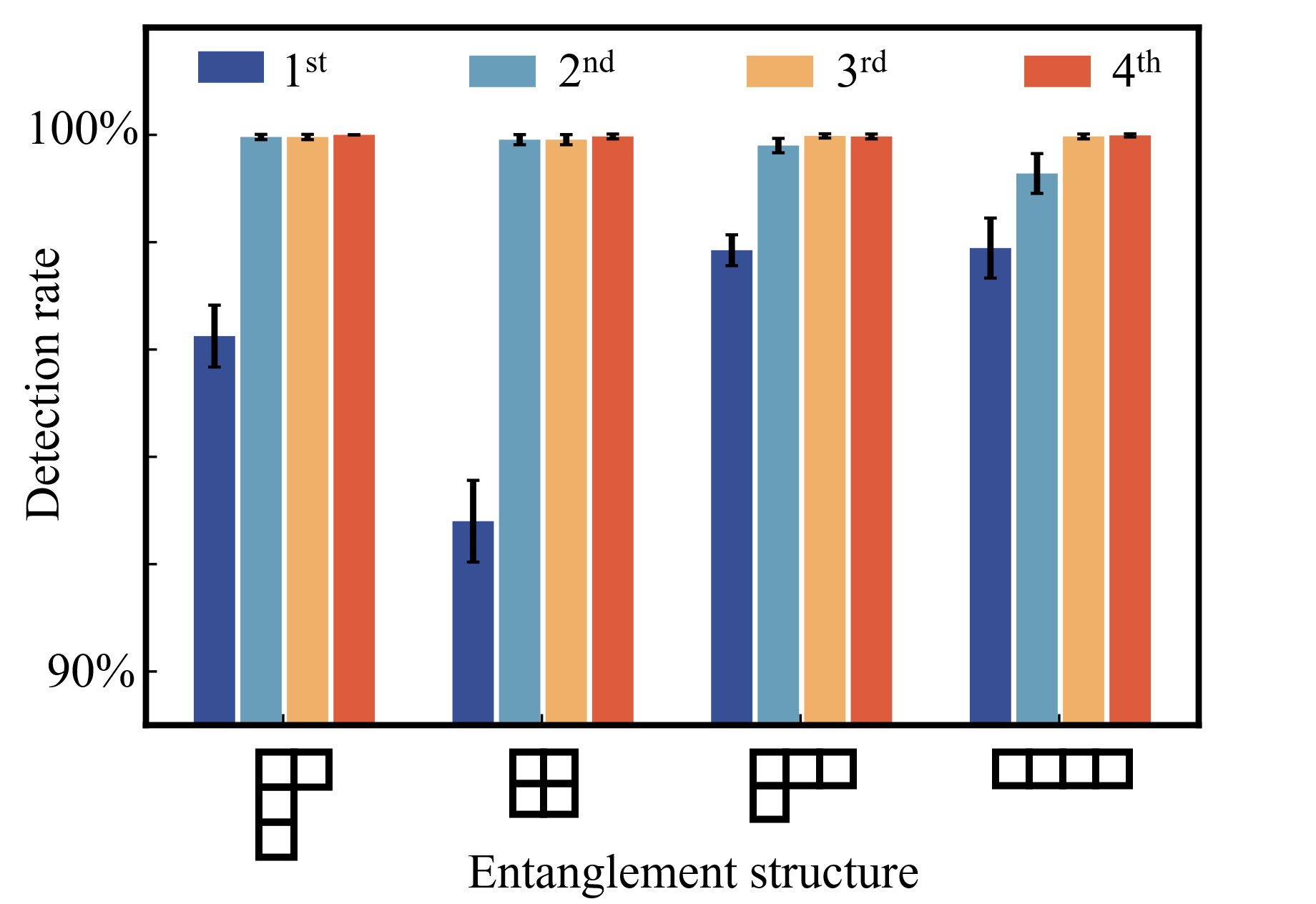}
    \caption{
    Multipartite entanglement structure detection.
   Without loss of generality, we randomly generated $10^4$ four-mode entangled states for each entanglement structure, classified according to Young diagram. The vertical axis shows the success rate of detecting this type of entanglement.
   The color denotes the order of the encoding operators $\mathbf{S}_j$ in QFI. The error bars represent the standard deviation of the detection rate, calculated by dividing these $10^4$ states into 10 sets.}
    \label{fig2}
\end{figure}

\vspace{2mm}
\textit{Characterizing entanglement structure of random Gaussian and non-Gaussian states.---}
The generation of a random CV quantum state $\hat{\rho}$ begins with a core state $|C\rangle$ using the stellar formalism~\cite{chabaud2020stellar, chabaud2021, chabaud2022holomorphic}. A multimode quantum state is then generated by applying an $m$-mode random Gaussian unitary operation $\hat{G}$ to the core state, namely as $|\psi\rangle = \hat{G}|C\rangle$. This procedure results in states that include the most common Gaussian and non-Gaussian states prepared in experiments, such as squeezed and photon subtracted states (see details in the Supplemental Material~\cite{supplementary}).
The entanglement structure of these randomly generated multimode states cannot be characterized by 
the previous criteria based on the covariance matrix~\cite{van-loock2003,threecolor,reidpra2014,duan2000,simon2000,Gessner2017entanglement,suxiaolong-npj2019}.  These methods are limited in that they either target a fixed partition or rely on second-moment measurements, making them suited only for Gaussian states. In contrast, our approach based on the QFI overcomes both limitations.

To test the effectiveness of our method, We randomly generate $40,000$ four-mode quantum states (ensuring 10,000 states for each type of entanglement structure by Peres–Horodecki criterion~\cite{peres1996-ppt,Horodecki1996-ppt,bound}). 
As shown in Fig.~\ref{fig2}, our method, based on QFI , demonstrates a high success rate in detecting entanglement structures.
For four modes, using first-order encoding operators $\hat{A}$ (combinations of $\hat{x}$ and $\hat{p}$) we can reach over 90\% detection rate.
Moreover, by including higher-order operators ($\hat{x}^2, \hat{p}^2, \hat{x}\hat{p} + \hat{p}\hat{x},\dots$), the success rate of characterizing entanglement structure increases to almost 100\%.

\begin{figure}[tb]
    \centering
    \includegraphics[width=8cm]{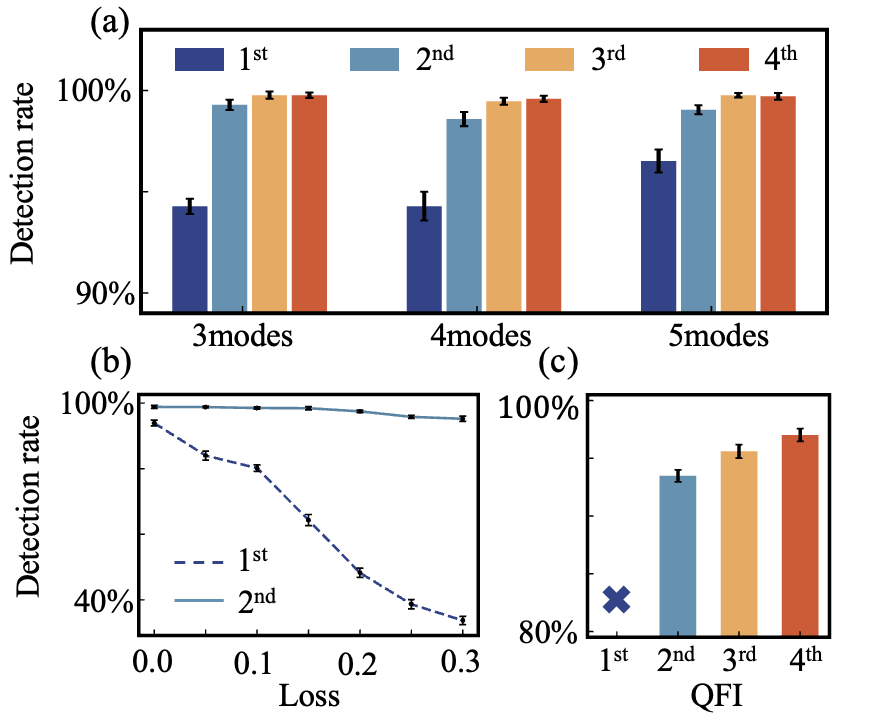}
    \caption{
     Witnessing fully inseparable entanglement for different non-Gaussian states.
(a)~The success rate of detecting entanglement for different multimode cases, where we randomly generated $10^4$ fully inseparable states of 3, 4, and 5 modes. The color denotes the order of the encoding operators in QFI. The error bars represent the standard deviation of the percentage of detecting fully inseparable states, calculated by dividing these $10^4$ states into 10 sets.
(b)~The percentage of detectable fully inseparable states in the presence of channel loss, where the dashed blue and solid green denote the QFI criterion based on first-order and second-order operators, respectively.
   (c)~The witness of full inseparability using QFI based on the three-photon SPDC process, where we generated $10^4$ states with random nonlinear strengths and plotted the percentage of detectable full inseparable states using operators of different orders.}
    \label{fig3}
\end{figure}

Additionally, the effectiveness of our method is not confined to four-mode systems, similar results can also be extended to multimode systems. As illustrated in Fig.~\ref{fig3}(a),  
our approach exhibits a high success rate in detecting fully inseparable states in three- and five-mode systems. 
 {
 In three-mode systems, by using first-order encoding operators $\hat{A}$ (linear combinations of $\hat{x}$ and $\hat{p}$), the success rate of detecting fully inseparable entanglement states is approximately 94.3\%.
}
By incorporating second-order encoding operators ($\hat{x}^2, \hat{p}^2, \hat{x}\hat{p} + \hat{p}\hat{x}$), the successful detecting rate  increases to 99.3\%. This performance further improves by extending the operator family to higher orders, achieving 99.8\% success rate with fourth-order operators.

Furthermore, we evaluate the performance of our method in the presence of a loss channel. Losses are introduced to the pure entangled state $ |\psi\rangle = \hat{G}|C\rangle $ using a single-mode loss channel $ \hat{L}_i(\eta) $ with efficiency coefficient $ \eta $, as detailed in Ref.~\cite{Eaton2022measurementbased}. 
The state with loss is described by:
$    
\hat{\rho} = \left(\prod_{i=1}^m \hat{L}_i(\eta_i)\right) \hat{G}|C\rangle\langle C| \hat{G}^{\dagger}\left(\prod_{i=1}^m \hat{L}_i^{\dagger}(\eta_i)\right).
$
Increased channel loss ($1-\eta$) impacts both the entanglement itself and the successful detection rate. As indicated by the dashed blue curve in Fig.~\ref{fig3}(b), the percentage of detectable fully inseparable states using first-order QFI decreases with increasing loss. Intriguingly, we demonstrate that extending the analysis to include second-order operators significantly enhances the successful detection rate, showing robustness against channel loss. This improvement is depicted in the solid green curves of Fig.~\ref{fig3}(b).

In addition to examining non-Gaussian states generated from core states, we also explore non-Gaussian states produced by nonlinear processes, such as the three-photon SPDC process. This process has been realized experimentally~\cite{spdc2004,prxthreemode} and has theoretically attracted significant interest due to its challenges in characterizing its correlations~\cite{kamle2018-prl,reid2015-prl-cat,agusta2020-prl-3spdc,zhangda-pra2021,zhangdaprl2021,tian2022,zhang2023-prl}.
Our method is not limited to these specific SPDC processes. To demonstrate its broad applicability, we randomly generated $10^4$ non-Gaussian states using the three-photon SPDC process, described by the Hamiltonian
$
    \hat{H} = (\chi_1 \hat{a}\hat{b}^2 + \chi_2 \hat{b}\hat{c}^2 + \chi_3 \hat{c}\hat{a}^2) + h.c.
$
Here, we evolve vacuum states for a duration $t$ with randomly selected parameters $\chi_i t$ from the interval [0, 0.04]. As illustrated in Fig.~\ref{fig3}(c), the specific form of the three-photon Hamiltonian means that the first-order QFI fails to detect any quantum correlations. In contrast, higher-order QFI successfully identifies entanglement. The reason the detection rate of fully inseparable states does not reach 100\% is that the random values for $\chi_i t$ have some probability of being very small, resulting in states with 
very weak multimode correlations.
These results exhibit the versatility of our method across different non-Gaussian states, demonstrating its effectiveness in accurately identifying operators that capture complex non-Gaussian correlations.

\textit{Conclusion.---} We provided a general method for characterizing entanglement for arbitrary Gaussian and non-Gaussian states. Based on the QFI, we established a systematic approach to analytically identify an operator to efficiently capture different quantum correlations. This method involves three key steps: (i) identifying convex sets of operators that can be used to witness $\mathcal{K}$-partition entanglement, (ii) determining the intersection of these sets, and (iii) selecting an operator to efficiently witness entanglement at the intersection.
To verify the accessibility and effectiveness of our method, we randomly generated $10^5$ states based on core states and nonlinear processes, which demonstrate a very high proportion of success in detecting entanglement structure. Furthermore, we exhibit the advantage of our method in resisting channel loss and detecting non-Gaussian entanglement. This approach provides a systematic way to distinguish entanglement structure for arbitrary states in CV systems.

Additionally, our method requires only one QFI encoding operator to capture the entanglement structure, making it more practical and experiment-friendly.
When applied to the nonlinear squeezing parameter~\cite{manuel2019-prl-qfi}, this operator enables efficient detection of entanglement via high-order quadratures~\cite{supplementary}, drastically reducing the number of observables needed compared with the full set of high-order quadratures.
Moreover, while high-order quadrature measurements remain more time-consuming than first-order ones, we notice that recent work has addressed this by developing a homodyne pattern-based detection method using neural networks~\cite{gao2024prl}. 
Building on our approach, they further constructed homodyne pattern-based strategies for multimode entanglement detection~\cite{gao2024}.
Together, these findings fill the gap for exploring and detecting complex entanglement structures in experiment-friendly ways.


\textit{Acknowledgements}---
This work is supported by the National Natural Science Foundation of China (No. 12125402, No. 12534016, and No. 12474256), the Innovation Program for Quantum Science and Technology (No. 2024ZD0302401 and No. 2021ZD0301500), and Beijing Natural Science Foundation (Grant No. Z240007). M.F. was supported by the Swiss National Science Foundation Ambizione Grant No. 208886, and by The Branco Weiss Fellowship -- Society in Science, administered by the ETH Z\"{u}rich. This work is supported by the project PID2023-152724NA-I00, with funding from MCIU/AEI/10.13039/501100011033 and FSE+, by the project CNS2024-154818 with funding by MICIU/AEI /10.13039/501100011033, by the project RYC2021-031094-I, with funding from MCIN/AEI/10.13039/501100011033 and the European Union ‘NextGenerationEU’ PRTR fund, by the project CIPROM/2022/66 with funding by the Generalitat Valenciana, and by the Ministry of Economic Affairs and Digital Transformation of the Spanish Government through the QUANTUM ENIA Project call—QUANTUM SPAIN Project, by the European Union through the Recovery, Transformation and Resilience Plan—NextGenerationEU within the framework of the Digital Spain 2026 Agenda, and by the CSIC Interdisciplinary Thematic Platform (PTI+) on Quantum Technologies (PTI-QTEP+). This work is supported through the project CEX2023-001292-S funded by MCIU/AEI.

\textit{Data availability}---The data that support the findings of
this article are openly available~\cite{mingshengtian2025rawdata}.
%

\end{document}


\title{Supplemental Material for ``Characterizing the Multipartite Entanglement Structure of Non-Gaussian Continuous-Variable States with a Single Evolution Operator''}
	
 \author{Mingsheng Tian}
 \thanks{These authors contributed equally to this work.}

 \address{State Key Laboratory for Mesoscopic Physics, School of Physics, Frontiers Science Center for Nano-optoelectronics, $\&$ Collaborative Innovation Center of Quantum Matter, Peking University, Beijing 100871, China}
  \address{Department of Physics, The Pennsylvania State University, University Park, Pennsylvania, 16802, USA}
 
 \author{Xiaoting Gao}
 \thanks{These authors contributed equally to this work.}
   \address{State Key Laboratory for Mesoscopic Physics, School of Physics, Frontiers Science Center for Nano-optoelectronics, $\&$ Collaborative Innovation Center of Quantum Matter, Peking University, Beijing 100871, China}

 \author{Boxuan Jing}
 \address{State Key Laboratory for Mesoscopic Physics, School of Physics, Frontiers Science Center for Nano-optoelectronics, $\&$ Collaborative Innovation Center of Quantum Matter, Peking University, Beijing 100871, China}

 \author{Fengxiao Sun}
   \address{State Key Laboratory for Mesoscopic Physics, School of Physics, Frontiers Science Center for Nano-optoelectronics, $\&$ Collaborative Innovation Center of Quantum Matter, Peking University, Beijing 100871, China}

\author{Matteo Fadel}
\affiliation{Department of Physics, ETH Z\"{urich}, 8093 Z\"{urich}, Switzerland}

\author{Manuel Gessner}
\affiliation{Departament de Física Teòrica, IFIC, Universitat de València, CSIC, C/ Dr. Moliner 50, 46100 Burjassot (València), Spain}

\author{Qiongyi He}
\email{qiongyihe@pku.edu.cn}
\address{State Key Laboratory for Mesoscopic Physics, School of Physics, Frontiers Science Center for Nano-optoelectronics, $\&$ Collaborative Innovation Center of Quantum Matter, Peking University, Beijing 100871, China}
\address{Collaborative Innovation Center of Extreme Optics, Shanxi University, Taiyuan, Shanxi 030006, China}
\address{Peking University Yangtze Delta Institute of Optoelectronics, Nantong, Jiangsu 226010, China}
\address{Hefei National Laboratory, Hefei 230088, China}

\maketitle
\tableofcontents

\section{Criterion based on Quantum Fisher Information}\label{qfi}

Originally, the Fisher information was introduced in the context of parameter estimation (see Ref.~\cite{qfi2018-rmp} for a review). To infer the value of $\theta$, one performs a measurement $\hat{M}=$ $\left\{\hat{M}_{\mu}\right\}$, which in the most general case is given by a positive operator valued measure (POVM). The Fisher information $F[\hat{\rho}(\theta), \hat{M}]$ quantifies the sensitivity of n independent measurements and gives a bound on the accuracy to determine $\theta$ as $(\Delta \theta)^{2} \geq 1 / (nF[\hat{\rho}(\theta), \hat{M}])$ in central limit ($n\gg1$). In particular, the Fisher information is defined as \cite{qfi1994}
\begin{equation}
    F[\hat{\rho}(\theta), \hat{M}] \equiv \sum_{\mu} \frac{1}{P(\mu \mid \theta)}\left(\frac{\partial P(\mu \mid \theta)}{\partial \theta}\right)^{2},
\end{equation}
where $P(\mu \mid \theta) \equiv \operatorname{Tr}\left\{\hat{\rho}(\theta) M_{\mu}\right\}$ is the probability to obtain the measurement outcome $\mu$ in a measurement of $\hat{M}$ given the state $\hat{\rho}(\theta).$

The Fisher information for an optimal measurement, i.e., the one that gives the best resolution to determine $\theta$, is called quantum Fisher information (QFI), and is defined as $F_{Q}[\hat{\rho}(\theta)] \equiv$ $\max _{\hat{M}} F[\hat{\rho}(\theta), \hat{M}] .$ There, one is interested in distinguishing the state $\hat{\rho}$ from the state $\hat{\rho}({\theta})=e^{-i \hat{A} \theta} \hat{\rho} e^{i \hat{A} \theta}$, obtained by applying a unitary induced by a Hermitian generator $\hat{A}$.  With the spectral decomposition $\hat{\rho}=\sum_k p_k|\Psi_k\rangle \langle \Psi_k|$, an explicit expression for $F_Q[\hat{\rho},\hat{A}]$ is given by \cite{qfi1994}
\begin{equation}
F_{Q}[\hat{\rho}, \hat{A}]=2 \sum_{\substack{k, l\\p_k+p_l\neq 0}} \frac{\left(p_{k}-p_{l}\right)^{2}}{p_{k}+p_{l}}\left|\left\langle\Psi_{k}\right|\hat{A}\left| \Psi_{l}\right\rangle\right|^{2}.
\end{equation}
And in pure states, it takes the simple form $ F_{Q}[|\psi\rangle\langle\psi|, \hat{A}]=4(\Delta A)^{2}.  $

The metrological witness for entanglement depends on the choice of the local operator $\hat{A}$. Certain choices of operators may be better suited than others to detect entanglement in a given state $\hat{\rho}$. In order to find an optimal operator, we can construct $\hat{A}_j$ by analytically optimizing over linear combinations of accessible operators from a set $\mathbf{S}_j = \{S_j^{(1)}, S_j^{(2)}, \cdots\}$:
\begin{equation}
    \hat{A}_j = \sum_{m=1}^{L} c_j^{(m)} \hat{S}_j^{(m)} = \mathbf{c}_j \cdot \mathbf{S}_j  \;,
\end{equation}
where the $\mathbf{c}_j=(c_j^{(1)},c_j^{(2)},\cdots,c_j^{(L)})$ are vectors of coefficients. A typical choice for $\mathbf{S}_j$ is 
$\mathbf{S}_j = \{\hat{x}_j, \hat{p}_j\}$, including local position operators and momentum operators.
However, when considering the non-Gaussian states, simply measuring linear observables is insufficient. We need to extend the $\mathbf{S}_j$ by incorporating higher-order moments of local quadrature operators. For example, extending to second-order quadratures, 
$
\{\hat{x}_j^2,\hat{p}_j^2,(\hat{x}_j\hat{p}_j+\hat{p}_j\hat{x}_j)/2
\}
$ 
should be added. And for third-order quadratures, 
$
\{\hat{x}_j^3,\hat{p}_j^3,\hat{x}_j\hat{p}_j^2,\hat{p}_j\hat{x}_j^2
\} 
$ should be included.
Extending to $m$th-order quadratures, we require $(m^2$$+$$3m)/2$ operators.
 {
With this,
the full operator
$
\hat{A}(\mathbf{c}) = \sum_{j=1}^{N} \hat{A}_j = \sum_{j=1}^{N} \mathbf{c}_j \cdot \mathbf{S}_j
$
is characterized by the combined vector 
$
\mathbf{c}=(\mathbf{c}_1,\cdots\mathbf{c}_N)^T
$
which contains $N\times L$ elements.
}
According to Eq.~(1) in main text, the quantity 
\begin{equation}
    W = F_Q[\hat{\rho}_{\mathcal{K}},   \hat{A}(\mathbf{c})] -
4 \mathcal{V}(\hat{\rho}_{\mathcal{K}},  \hat{A(\mathbf{c})})
\end{equation}
must be nonpositive for arbitrary choices of $\mathbf{c}$ whenever the state $\hat{\rho}$ is separable. And the witness of $W>0$ can certify entanglement. Thus we can maximize $W[\hat{\rho},\hat{A}(\mathbf{c})]$ by variation of $\mathbf{c}$ to obtain an optimized entanglement witness for the state $\hat{\rho}$.

To this aim, let us first express the quantum Fisher information in matrix form as $F_Q[\hat{\rho},\hat{A}(\mathbf{c})]=\mathbf{c}^T Q_{\hat{\rho}}^{\mathcal{ {S}}}\mathbf{c}$, 
 {
where $Q_{\hat{\rho}}^{\mathcal{S}}$ is an $NL\times NL$ matrix, with elements given by
}
$$
(Q_{\hat{\rho}}^{\mathcal{ {S}}})_{i j}^{m n}=2 \sum_{k, l} \frac{(p_{k}-p_{l})^{2}}{p_{k}+p_{l}}\langle\Psi_{k}|\hat{ {S}}_{i}^{(m)}| \Psi_{l}\rangle\langle\Psi_{l}|\hat{ {S}}_{j}^{(n)}| \Psi_{k}\rangle.
$$
Here, the spectral decomposition $\hat{\rho}=\sum_k p_k|\Psi_k\rangle\langle\Psi_k|$ defines
element-wise and the sum extends over all pairs with $p_k+p_l\neq 0$. The indices $i$ and $j$ refer to different parties, while the indices $m$ and $n$ label the respective local sets of observables. 
 {
The indices $(i$$-$$1)L $$+$$ m$ and $(j$$-$$1)L $$+$$ n$ specify the unique row and column positions of the matrix $Q_{\hat{\rho}}^{\mathcal{S}}$.
}
Similarly, we can express the  elements of the covariance matrix of $\hat{\rho}$ as $(\Gamma_{\hat{\rho}}^{\mathcal{ {S}}})_{i j}^{m n}=\mathrm{Cov}(\hat{ {S}}_{i}^{(m)}, \hat{ {S}}_{j}^{(n)})_{\hat{\rho}}$. If the above covariance matrix is evaluated after replacing $\hat{\rho}$ with $\Pi(\hat{\rho})=\hat{\rho}_{1} \otimes \cdots \otimes \hat{\rho}_{N}$, where $\hat{\rho}_i$ is the reduced density operator, we arrive at the expression for the local variances, $\sum_{j=1}^{N} \mathrm{Var}(\mathbf{c}_{j}\cdot \hat{\mathbf{A}}_{j})_{\hat{\rho}}=\mathbf{c}^{T} \Gamma_{\Pi(\hat{\rho})}^{\mathcal{A}} \mathbf{c}$. Combining this with expression for the quantum Fisher matrix, the separability criterion reads 
\begin{equation}\label{eqW1}
W[\hat{\rho}, \hat{A}(\mathbf{c})]=
\frac
{\mathbf{c}^{T}\left(Q_{\hat{\rho}}^{\mathcal{ {S}}}-4 \Gamma_{\Pi(\hat{\rho})}^{\mathcal{ {S}}}\right) \mathbf{c} }
{\mathbf{c}^T \mathbf{c}}
\leq0.
\end{equation}
An entanglement witness is therefore found when the matrix $ {\mathcal{M}}=(Q_{\hat{\rho}}^{\mathcal{ {S}}}-4 \Gamma_{\Pi(\hat{\rho})}^{\mathcal{ {S}}})$ has at least one positive eigenvalue. 
For pure states $\hat{\rho}=|\Psi\rangle\langle\Psi|$, the quantum Fisher matrix coincides, up to a factor of 4, with the covariance matrix, i.e., $Q_{|\Psi\rangle}^{\mathcal{ {S}}}=4 \Gamma_{|\Psi\rangle}^{\mathcal{ {S}}}$. Thus, 
the matrix $ {\mathcal{M}}$ can be simplified to
$  {\mathcal{M}}=\Gamma_{|\Psi\rangle}^{\mathcal{ {S}}}-\Gamma_{\Pi(\mid \Psi))}^{\mathcal{ {S}}} $.
%

\section{Generation of random Gaussian and non-Gaussian states}\label{sec2}

To generate random Gaussian and non-Gaussian states, we first characterize them using the stellar formalism~\cite{chabaud2020stellar,chabaud2021,chabaud2022holomorphic}. 
In this formalism, we analyze an $m$-mode pure state $|\psi\rangle$ in terms of its stellar function $F_\psi^{\star}(\mathbf{z})$. 
To define this function, we start by considering the state's decomposition in the Fock basis, i.e., $|\psi\rangle=\sum\limits_{\mathbf{n}\geq0}\psi_{\mathbf{n}}|\mathbf{n}\rangle\in\mathcal{H}^{\otimes m}$ with $\mathbf{n}=(n_1,n_2,\cdots,n_m)$, such that the stellar function can be written as
\begin{align}
F_\psi^{\star}(\mathbf{z}) \equiv e^{\frac{1}{2}\|\mathbf{z}\|^2}\left\langle \mathbf{z}^*|\psi\right\rangle=\sum\limits_{n_1,n_2,\cdots n_m} \frac{\psi_\mathbf{n}}{\sqrt{n_1!n_2!\cdots n_m!}} z_1^{n_1}z_2^{n_2}\cdots z_2^{n_m},~~\forall~\mathbf{z}=(z_1,z_2,\cdots,z_m)\in\mathbb{C}^2
\end{align}
where $|\mathbf{z}\rangle=\text{e}^{-\frac{1}{2}\|\mathbf{z}\|^2}\sum\limits_{n_1,n_2,\cdots,n_m}\frac{z_1^{n_1}z_2^{n_2}\cdots z_m^{n_m}}{\sqrt{n_1!n_2!\cdots n_m!}}|n_1,n_2,\cdots,n_m\rangle$ is an $m$-mode coherent state of complex amplitude $\mathbf{z}$. 
The stellar rank $r$ of $|\psi\rangle$ is defined as the number of zeros of its stellar function, representing a minimal non-Gaussian operational cost to engineer the state from the vacuum. 
For instance, $r=0$ means that the state is Gaussian, while $r=1$ corresponds to a class of non-Gaussian states that contains, both, single-photon added and subtracted states~\cite{chabaud2022holomorphic}. Any multimode pure state $|\psi\rangle$ with finite stellar rank $r$ can be decomposed into $|\psi\rangle=\hat G|C\rangle$, where $\hat G$ is a Gaussian operator acting onto the state $|C\rangle$, which is called core state; it is a normalized pure quantum state with multivariate polynomial stellar function of degree $r$, equal to the stellar rank of the state. 
It then follows immediately that Gaussian operations $\hat G$ must preserve the stellar rank~\cite{chabaud2021}. \par

\begin{figure}
    \centering
    \includegraphics[width=1\linewidth]{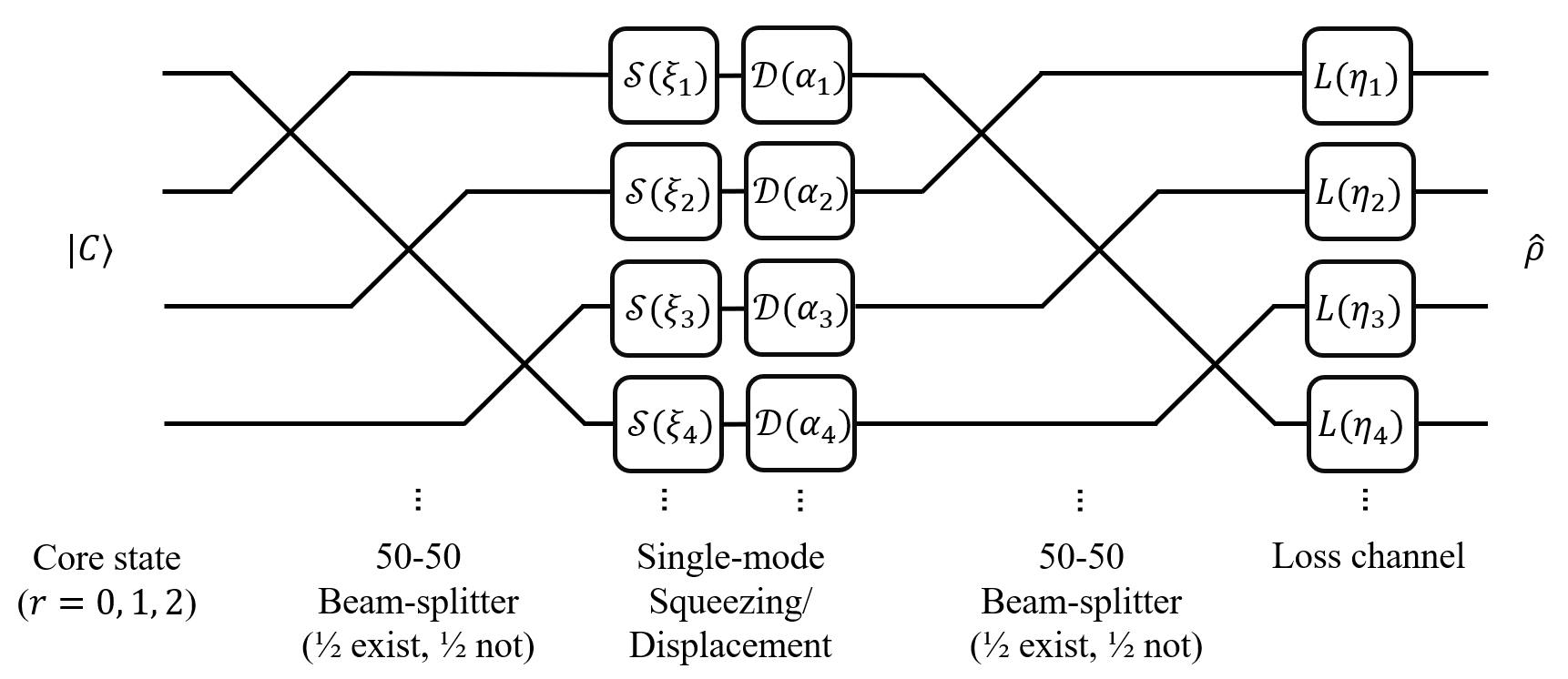}
    \caption{The generation process of multimode random Gaussian and non-Gaussian states.}
    \label{sfig1}
\end{figure}

Our randomly generated non-Gaussian states follow the above decomposition, which begins with a core state $|C\rangle$ with a given stellar rank $r$ and random complex superposition coefficients of Fock basis. A multimode core state $|C\rangle$ is a finite superposition of multimode Fock states, whose stellar rank corresponds to the minimal number of photon additions that are necessary to engineer the state from the vacuum~\cite{chabaud2021}.
We restrict $r$ to  $0$, $1$ and $2$, which includes the most common Gaussian and non-Gaussian states in experiments. 
According to the Williamson decomposition and the Bloch-Messiah decomposition, an $m$-mode Gaussian unitary operation $\hat G$ can be decomposed as $\hat G=\mathcal{\hat U}(\varphi)\left( \prod\limits_{i=1}^m \mathcal{\hat S}_i(\xi_i)\mathcal{\hat D}_i(\alpha_i)\right)\mathcal{\hat V}(\phi)$, where $\mathcal{\hat S}_i(\xi_i)=\text{e}^{\frac{1}{2}(\xi_i^*\hat a_i^2-\xi_i\hat a_i^{\dagger2})}$ is a squeezing operator with complex squeezing parameter $\xi_i$ acting on mode $i$ and $\mathcal{\hat D}_i(\alpha_i)=\text{e}^{\alpha_i\hat a^\dagger_i-\alpha_i^*\hat a_i}$ is a displacement operator with complex displacement amplitude $\alpha_i$ acting on mode $i$.  
$\mathcal{\hat U}(\varphi)$ and $\mathcal{\hat V}(\phi)$ are passive Gaussian transformations with complex coupling $\varphi$ and $\phi$, which are set to exist in the circuit with a $50\%$ probability.

Losses are also added to each mode of the pure state
$|\psi\rangle=\hat G|C\rangle$
, using a single-mode loss channel $\hat L_i(\eta_i)$ with efficiency coefficient $\eta_i$ as described in Ref.~\cite{Eaton2022measurementbased}. 
The entire quantum circuit is shown in Fig.~\ref{sfig1}, generating an $m$-mode state $\hat\rho$ with several randomly selected free parameters, $\mathbf{\xi}$ and $\mathbf{\alpha}\in [0,0.05]$, given by
\begin{align}
\hat\rho=\left(\prod_{i=1}^m\hat L_i(\eta_i)\right)\hat G |C\rangle\langle C|\hat G^\dagger \left(\prod_{i=1}^m\hat L_i^\dagger(\eta_i)\right).
\end{align}\par

\section{The examples for characterizing multimode quantum states}
To evaluate the efficacy of our method for characterizing entanglement structures, we generate random entangled states according to different entanglement structure, as described in Sec.~\ref{sec2}. For example, a four-mode entangled state with a (2,1,1) structure can be created by taking the direct product of a random two-mode state and two single-mode states. However, since these randomly generated states may not always be fully entangled, it remains unclear whether a state's non-detection by our method is due to its limitations or the absence of entanglement in the state itself. To address this, we establish a benchmark to filter out non-entangled states, thereby refining the set of states used to assess our method.
Here, we can apply the Peres-Horodecki criterion~\cite{peres1996-ppt,Horodecki1996-ppt}, also known as the Positive Partial Transpose (PPT) criterion, to filter out non-entangled states. The reason we choose PPT criterion is that it is a robust tool for identifying entanglement, only weakly entangled states, known as bound entanglement states~\cite{bound}, cannot be detected by PPT criterion. 

\begin{figure}
    \centering
    \includegraphics[width=15cm]{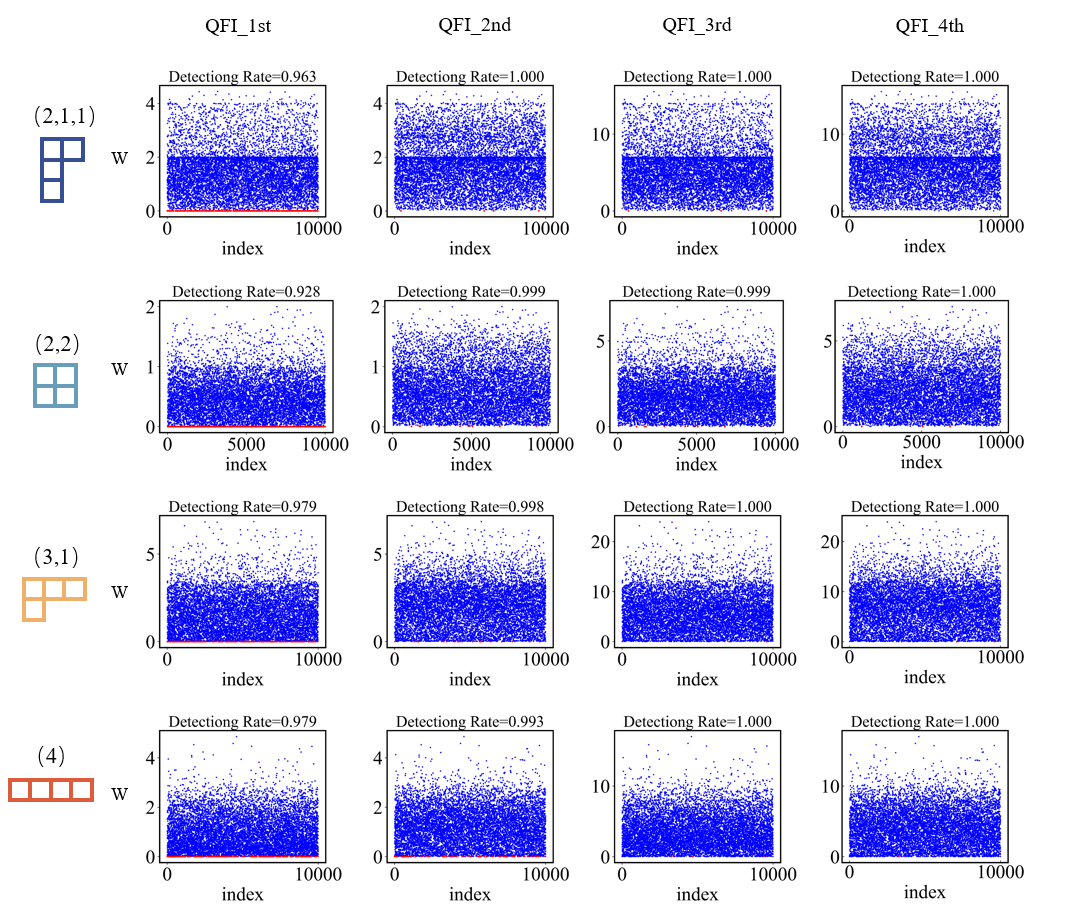}
    \caption{The witness of four-mode entanglement structure (see also Fig.~1 in the main text). We generate $10^4$ random entangled states for each case, with the horizontal axis representing the index of entangled states and the vertical axis representing the entanglement witness. States with $ W > 0 $ (indicating entanglement) are marked in blue, while the red point indicates the states cannot be verified entanglement.
    }
    \label{sfig5}
\end{figure}

We generate a large sample of states, excluding those that fail to exhibit entanglement according to the PPT criterion. From this, we select 40,000 entangled states—ensuring 10,000 states for each type of entanglement structure—to test the effectiveness of our method.
As depicted in Fig.~\ref{sfig5}, our detailed analysis involves calculating the value of $ W $ for 10,000 different entangled states within each category of Young-diagrams type entanglement. A positive $ W $ value confirms the presence of entanglement. The results showcase a high detection rate across all types of entanglement, validating the robustness of our method in identifying diverse entangled states.

Multipartite entanglement is an essential resource for many quantum information and computation protocols.
However, traditional criteria~\cite{van-loock2003}, which rely on the variances of linear combinations of position and momentum operators cannot effectively capture multimode entanglement structure for two primary reasons: first, they only consider linear combinations of first-order position and momentum operators, which are insufficient for detecting higher-order correlations indicative of non-Gaussian entanglement; second, the complexity of multimode systems complicates the identification of effective combinations of position and momentum operators to witness  entanglement. 

\begin{figure}
    \centering
    \includegraphics[width=1\linewidth]{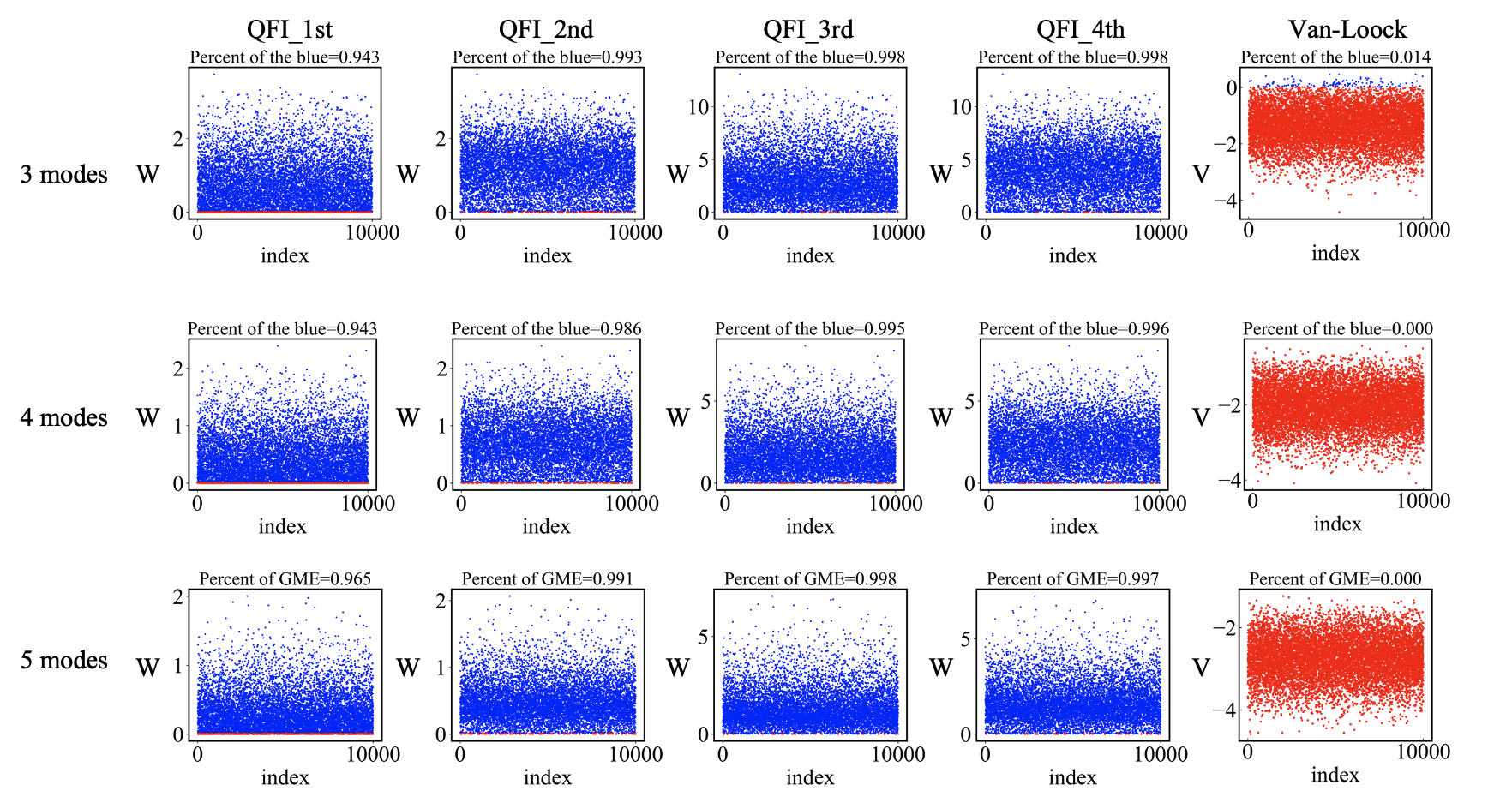}
    \caption{The witness of full inseparability for random non-Gaussian entangled states with stellar rank $r=2$ (see also Fig.~2 (a) in the main text). The horizontal axis represents the index of entangled states and the vertical axis represents the entanglement witness. States with $ W (V) > 0 $ (indicating full inseparability) are marked in blue, while states that cannot be detected as fully inseparable states are marked in red. The Van-Loock criterion fails to capture the Non-Gaussian entanglement.} 
    \label{sfig2}
\end{figure}

Based on the method mentioned in Sec.~\ref{sec2}, we can generate $10^4$ three, four, and five fully inseparable non-Gaussian states with stellar $r=2$. As shown in Fig.~\ref{sfig2}, our method based on QFI exhibits good performance in detecting entanglement.
In contrast, we also apply variance-based criteria~\cite{van-loock2003} for entanglement witness, where a separable state follows:
\begin{equation}
\left\langle (\Delta \hat{u})^2 \right\rangle_{\hat{\rho}} + \left\langle (\Delta \hat{v})^2 \right\rangle_{\hat{\rho}} \geqslant f(h_1, h_2, \ldots, h_N, g_1, g_2, \ldots, g_N),
\end{equation}
with:
\begin{equation}
\begin{aligned}
& \hat{u} \equiv h_1 \hat{x}_1 + h_2 \hat{x}_2 + \cdots + h_N \hat{x}_N, \\
& \hat{v} \equiv g_1 \hat{p}_1 + g_2 \hat{p}_2 + \cdots + g_N \hat{p}_N,
\end{aligned}
\end{equation}
Here, $ x_i $ and $ p_i $ represent the position and momentum operators, respectively, and $ h_i $ and $ g_i $ are their coefficients. The function $ f(\cdot) $ relates to these coefficients. Here we take $ g_1=h_1=1 $, $ g_k=-h_k=-1/\sqrt{N-1} $, and $ f(h_1, h_2, \ldots, h_N, g_1, g_2, \ldots, g_N)=1/(N-1) $.

\begin{figure}
    \centering
    \includegraphics[width=1\linewidth]{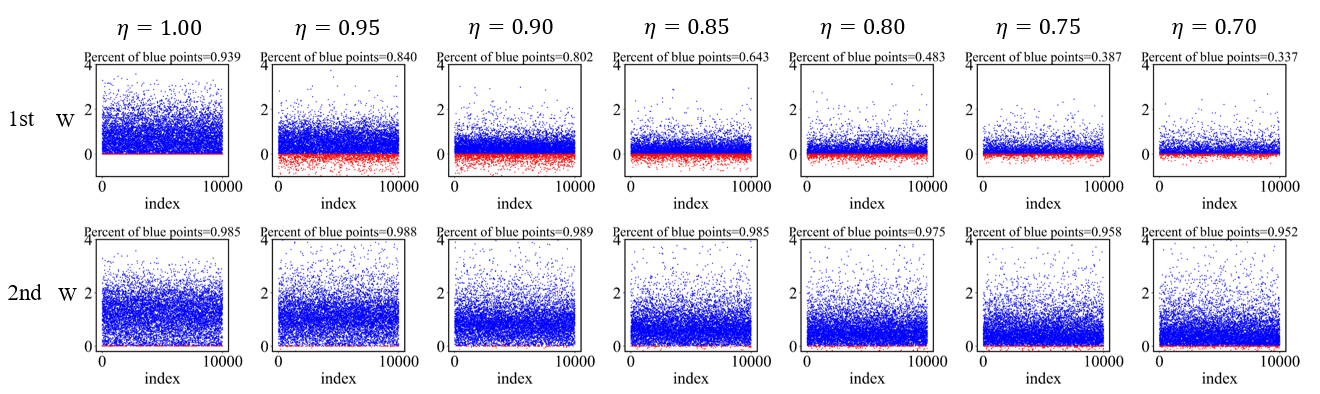}
    \caption{The witness of three-modes entanglement with the existence of loss by using first- and second-order operators (see also Fig.~2 (b) in the main text). Here, the horizontal axis represents the index of entangled states and the vertical axis represents the entanglement witness. States with full inseparability are marked in blue, while states that cannot be detected as fully inseparable states are marked in red.}
    \label{sfig3}
\end{figure}

For comparison purposes, we redefine the entanglement parameter $ V $ as:
\begin{equation}\label{seq:vanlock}
    V = \frac{4|\langle{[\hat{x_i},\hat{p_i}]}\rangle|}{N-1} - \left\langle (\Delta \hat{u})^2 \right\rangle_{\hat{\rho}} - \left\langle (\Delta \hat{v})^2 \right\rangle_{\hat{\rho}}.
\end{equation}
States with $ V > 0 $ indicate full inseparability. According to Eq.~(\ref{seq:vanlock}), we calculated the entanglement witness $ V $, where the states are the same as those detected by the QFI criterion. As shown in the far right plots of Fig.~\ref{sfig2}, it is not an efficient criterion to witness entanglement. 
\ms{
It is worth noting that the criterion used for comparison here is not optimal; better performance could, in principle, be achieved by employing suitable linear combinations. However, this is highly challenging due to the complexity of the entanglement structure and the exponential scaling of the parameter space with both the number of modes and the order of quadrature correlations (leading to a size of $(2M+1)^N$). Even for Gaussian states, identifying effective entanglement structure witnesses becomes increasingly difficult as the system size grows. For these reasons, simple linear witnesses are used as a comparison here.
}

Furthermore, we evaluate the performance of our method in the presence of a loss channel. Losses are introduced to the pure fully inseparable state $ |\psi\rangle = \hat{G}|C\rangle $ using a single-mode loss channel $ \hat{L}_i(\eta) $ with efficiency coefficient $ \eta $, as detailed in Ref.~\cite{Eaton2022measurementbased}. 
The state with loss is described by:
$   
\hat{\rho} = \left(\prod_{i=1}^m \hat{L}_i(\eta_i)\right) \hat{G}|C\rangle\langle C| \hat{G}^{\dagger}\left(\prod_{i=1}^m \hat{L}_i^{\dagger}(\eta_i)\right).
$
Increased channel loss ($1-\eta$) impacts both the entanglement itself and the detection of entanglement. As indicated in Fig.~\ref{sfig3}, the percentage of entanglement detectable using first-order QFI decreases with increasing loss. Intriguingly, we demonstrate that extending the analysis to include second-order operators significantly enhances the successful detection rate of entanglement, showing robustness against channel loss.

\begin{figure}
    \centering
    \includegraphics[width=1\linewidth]{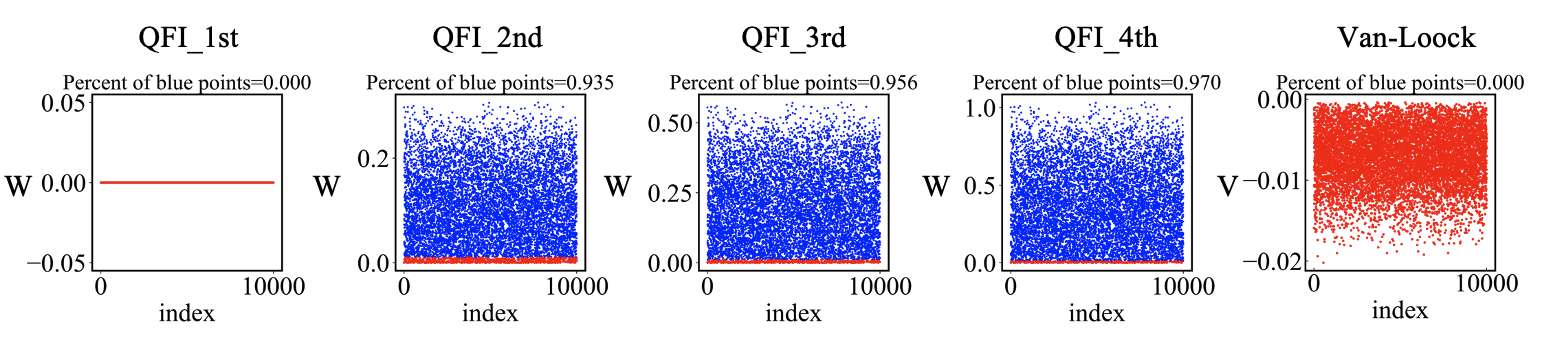}
    \caption{The witness of full inseparability for the states generated by SPDC process (see also Fig.~2 (c) in the main text). The horizontal axis represents the index of entangled states and the vertical axis represents the entanglement witness. States with full inseparability are marked in blue, while states that cannot be detected as fully inseparable states are marked in red.}
    \label{sfig4}
\end{figure}

In addition to examining non-Gaussian states generated from core states, we also explore non-Gaussian states produced by nonlinear processes, such as the three-photon spontaneous parametric down-conversion (SPDC) process. This process has been realized experimentally~\cite{spdc2004,prxthreemode} and has theoretically attracted significant interest due to its challenges for characterizing its correlations~\cite{kamle2018-prl,reid2015-prl-cat,agusta2020-prl-3spdc,zhangda-pra2021,zhangdaprl2021,zhang2023-prl}.
For our analysis, we randomly generated $10^4$ non-Gaussian states using the three-photon SPDC process, described by the following Hamiltonian:
\begin{equation}
    \hat{H} = (\chi_1 \hat{a}\hat{b}^2 + \chi_2 \hat{b}\hat{c}^2 + \chi_3 \hat{c}\hat{a}^2) + h.c.
\end{equation}
Here, the parameters $\chi_i t$ are randomly selected from the interval [0, 0.04], and the initial states are vacuum states. As illustrated in Fig.~\ref{sfig4}, the specific form of the three-photon Hamiltonian means that the first-order QFI fails to detect any quantum correlations. In contrast, higher-order QFI successfully identifies three-mode entanglement.
These results exhibit the versatility of our method across different non-Gaussian states, demonstrating its effectiveness in accurately identifying operators that capture complex non-Gaussian correlations.

\ms{
At last, we provide a concrete example to illustrate how our method significantly reduces the number of required observables compared to the whole covariance matrix or high-order moment tensor.
For an $N$-mode continuous-variable (CV) system with quadrature operators up to the $M$th order, the second-order covariance matrix extends to a high-order moment tensor, written as
\begin{equation}
    \mathcal{M}_{ij\cdots kl} = 
\mathbb{E}\left[(X_i - \overline{X}_i)(X_j - \overline{X}_j)
\cdots
(X_k - \overline{X}_k)(X_l - \overline{X}_l)\right],
\end{equation}
where $\mathbb{E}$ indicates the expectancy of observables.
The high-order moment tensor has a dimension of $L^M$, where $L=2N+1$ corresponds to the size of the local operator set $\{\mathbb{I},\hat{x}_i,\hat{p}_i,\hat{x}_j,\hat{p}_j,\cdots\
\hat{x}_k,\hat{p}_k,\hat{x}_l,\hat{p}_l\}$.
As both the mode number $N$ and quadrature order $M$ increase, the size of the moment tensor grows rapidly, making it challenging to find suitable observables to witness entanglement by using the conventional approach~\cite{van-loock2003}.
In contrast, our method enables the characterization of multipartite entanglement structure by only a small set of well-optimized observables, which are obtained by solving an eigenvalue problem for an $M^2N \times M^2N$ matrix. This optimization is computationally tractable, as it can be performed in guaranteed polynomial time using standard linear algebra techniques.

As an example, we consider a three-mode SPDC system 
i.e., $
\hat{H} = i\hbar \kappa ( \hat{b}\hat{a}_1^{\dagger} \hat{a}_2^{\dagger} \hat{a}_3^{\dagger} - \hat{b}^{\dagger} \hat{a}_1 \hat{a}_2 \hat{a}_3 )
$, where a pump photon of frequency $\omega_p$ is down-converted into three nondegenerate photons at frequencies $\omega_1, \omega_2, \omega_3$.
By extending the operator set to include second-order quadratures, an optimal generator can be constructed as $\hat{A}_{\text{opt}} = \sum_{j=1}^3 (\hat{x}_j^2 + \hat{p}_j^2)$. To detect multipartite entanglement in this system, we use the nonlinear squeezing parameter~\cite{manuel2019-prl-qfi}, which is equal to QFI after optimization. For the three-mode SPDC state, the optimized nonlinear squeezing parameter takes the form~\cite{tian2022}:
\begin{equation}
\begin{aligned}
\chi^{-2}
=\frac{|\langle
\hat{p}_1\hat{x}_2\hat{x}_3+\hat{x}_1\hat{p}_2\hat{x}_3+\hat{x}_1\hat{x}_2\hat{p}_3
\rangle_{\hat{\rho}}|^{2}}
{\Delta^2(\hat{x}_1\hat{x}_2\hat{x}_3)_{\hat{\rho}}}.
\end{aligned}
\end{equation}
This witness involves only four specific third- or sixth-order moment observables, in stark contrast to the full moment tensor (six-order). Hence, our method largely reduces the required observables for entanglement detection.
}

%
